# Is Basketball 'a Game of Runs?'

Mark F. Schilling
California State University Northridge

## Abstract

Basketball is often referred to as "a game of runs." We investigate the appropriateness of this claim using data from the full NBA 2016-17 season, comparing actual longest runs of scoring events to what long run theory predicts under the assumption that team "momentum" is not present. We provide several different variations of the analysis. Our results consistently indicate that the lengths of longest runs in NBA games are no longer than those that would occur naturally when scoring events are generated by a random process, rather than one that is influenced by "momentum".

# Introduction

On the evening of February 5, 1987, the Los Angeles Lakers' basketball team scored 29 consecutive points against the unfortunate Sacramento Kings, representing possibly the longest run of consecutive points in the history of U.S. professional basketball. While this event was exceptional to say the least, viewers of professional and college basketball games frequently witness runs by one team or the other, and television and radio announcers and commentators often remark that "Basketball is a game of runs." But is such a sweeping statement justified?

In order to answer this question, we first have to agree on what it means. Most people would concur that the implication of the phrase is that a team that has scored several times in a row has "momentum," giving them a higher chance of making the next score than when they are not on a run. This is a team level version of the well-known "hot hand" concept, which has been extensively studied at the player level in numerous sports and games (see for example Gilovich, Vallone and Tversky, 1985; Albright, 1993; Stern, 1997; Vergin, 2000; Ayton and Fischer, 2004; Crosson and Sundali, 2005; and Miller and Sanjurjo, 2018).

In the following we explore the question of whether "basketball is a game of runs" for professional basketball teams in the National Basketball Association (NBA), using an analysis of data from the entire NBA 2016-17 season. We frame the issue along the lines of a hypothesis test: Our null hypothesis is that if a team has just scored one or more times in a row, this does not affect their chances of being the next team to score; the alternative hypothesis is that the team's chance of scoring next increases. In order to make our inquiry manageable, we define a run in one of two ways: (i) as a set of consecutive field goals scored by one team, without the other team scoring a field goal; (ii) as a set of consecutive scoring events by one team, each event being either a field goal or one or more made free throws.

# Modeling NBA Scoring

In studies of whether an individual player has a hot hand, the usual null model is that the player's shot attempts are independent Bernoulli trials. Let's see what sort of runs an analogous model might lead to for team scoring in NBA basketball. Initially we will treat each team as having approximately equal chances of scoring when in possession of the ball. This is a fairly reasonable simplifying assumption, as differences in team abilities should balance out when analyzing the entire data set.

For now we will ignore free throws and consider only field goals, not distinguishing between two-point and three-point field goals. Using the website http://shiny.calpoly.sh/Longest_Run/ we can easily produce a random sequence of outcomes that simulates the scores by the two teams.

Here is a sequence of 80 such outcomes generated from this website, which corresponds to the approximate average number of field goals made in an NBA game during the 2016-17 season:

<div align="center">

A B B A B A A A B A A A B A A B B B B B B B A A B

A B A A B A B B B B B A B B A A B A B B A A A A A A B B

B A B B A B A B A B B A B B A B B B B B A A A B B B

</div>

Notice that there are several long runs of field goals, including runs of seven and five by Team B and a run of six by team A. This is a typical pattern of runs for a sequence of 80 Bernoulli trials with equal probabilities for A and B on each trial. With this model, long runs of scores by one team are a natural and frequent occurrence, without the need for "momentum" to explain them.

However, the simple independent Bernoulli trials model is not appropriate for basketball team scoring because we need to account for the fact that, by rule, the team that was just scored on generally receives possession of the ball. If one team scores a basket, then the other team inbounds the ball, receiving the next opportunity to score—thus the potential team scoring events in a basketball game are not independent. If one team has just scored, it is *less* likely that that same team will be the next to score. Therefore, the runs that occur in a basketball game would tend to be somewhat shorter than those produced by the independent Bernoulli trials model—unless basketball really is a "game of runs."

We can deal with the additional complication that this dependency produces by simply writing the data differently. The above sequence of **A**'s and **B**'s yields the derived sequence

$$\textbf{D S D D D S S D D S D D D D S D S S S S S S D} \;\ldots\;,$$

where the symbols D and S indicate whether a basket was scored by a **D**ifferent/the **S**ame team as the one that scored the previous basket. The run of six **S**'s seen here corresponds to the run of seven **B**'s in the simulated sequence of baskets above. In general, a run of $k$ consecutive **S**'s indicates a run of $k + 1$ consecutive scores by one team.

If our null hypothesis of no momentum is true, then the sequence of **D**'s and **S**'s generated in a game will be statistically independent. To judge how well this null model explains the observed patterns of scores in NBA basketball, we can compare the lengths of the longest runs of scores in each of the 1,230 games in the 2016-17 season to what is predicted by probability theory (Schilling, 1990; 2012; Das Gupta 2018; see the Appendix).

The theoretical probability distribution of the longest run of field goals by either Team A or Team B distribution depends on (i) the actual number of baskets scored in the game and (ii) $P(S)$, the probability that the same team that just scored scores the next basket. Using the data from all 1,230 NBA games in 2016-17, $P(S)$ is estimated by the sample proportion as 0.38.

Table 1 shows the resulting expected length of the longest run of field goals that will occur in a game, for varying numbers of total field goals scored by the two teams combined:

### Table 1. Expected Length of the Longest Run of Field Goals Under the Null Model

| Total field goals | Expected longest run length |
|---|---|
| 50 | 4.66 |
| 75 | 5.08 |
| 100 | 5.38 |
| 125 | 5.61 |
| 150 | 5.80 |

The expected longest run length grows approximately logarithmically with the number of trials (Schilling, 1990; 2012), thus its value does not vary much as the number of total scores in Table 1 changes. During the 2016-17 season, the total number of field goals scored ranged from a low of 52 to a high of 106. Thus in the absence of momentum a typical NBA game should have had a longest run of about five baskets in a row.

## What the Data Reveals

We performed several analyses of scoring patterns in order to look for evidence that the runs of scores in NBA basketball are longer than what probability theory predicts in the absence of "momentum." Averaging the expected longest run lengths over all NBA games in the 2016-17 season produces a theoretical season average longest run length of 5.12 field goals per game. Averaging the actual longest run lengths gives the nearly identical value 5.17. The average length of the longest run of baskets is only 1% longer than theory predicts, a difference well within the limits of chance variation.

While there is almost no difference in average and expected longest run length, there might be other differences between the distributions of theoretical and actual longest run lengths. We therefore conducted several more thorough analyses that compare the expected number of longest runs of each length to the actual frequency distribution of the longest runs in the totality of games played in 2016-17.

## Analysis of Field Goal Runs Distribution

Figure 1 compares actual and theoretical distributions for the longest run of field goals in an NBA game:

**Figure 1. Distribution of the Longest Run of Field Goals in Games: Predicted vs. Observed**

| Run Length | #Predicted | #Observed |
|---|---|---|
| 3 | 65.2 | 51 |
| 4 | 370.8 | 357 |
| 5 | 406.0 | 414 |
| 6 | 226.5 | 243 |
| 7 | 98.3 | 100 |
| 8 | 39.0 | 46 |
| 9 | 15.0 | 11 |
| 10 | 5.7 | 2 |
| 11 | 2.1 | 4 |
| 12 | 0.8 | 2 |
| Mean | 5.12 | 5.17 |

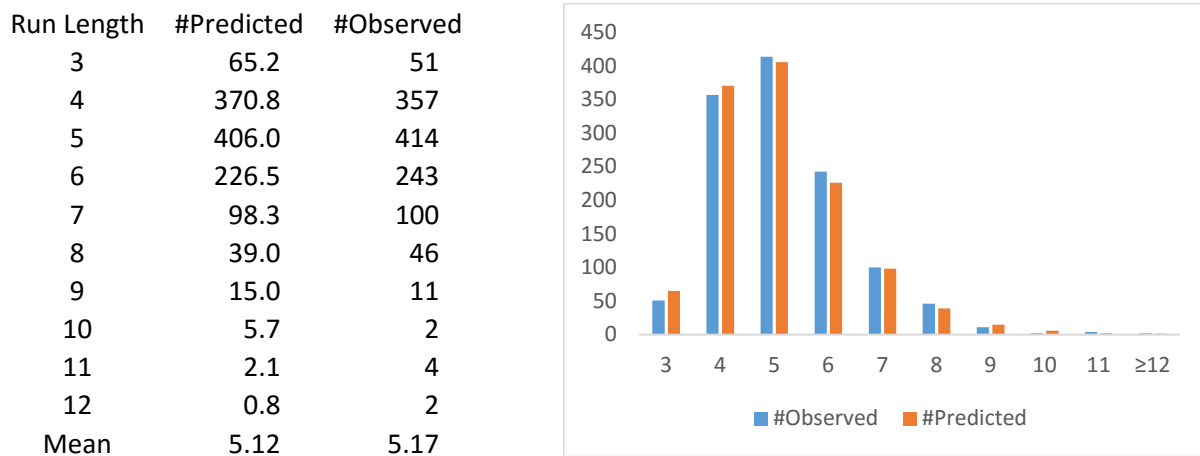

The fit of the null model to the data is excellent. A chi-square goodness of fit test gives a nonsignificant $p$-value of $p = 0.38$, confirming this assertion. Thus the distribution of actual longest run lengths is consistent with our null hypothesis of no momentum.

As a side note, one might conjecture that due to the advantage of having home court, the home team would achieve longer longest runs than the road team. The difference was slight, however, with a mean longest run length of 4.55 for home teams and 4.40 for road teams.

# Analysis Performed Separately for Each Half

If momentum is present in basketball games, we might doubt that it persists across halftime. Therefore a second analysis was performed in which each half was treated as a separate game. Figure 2 show the results:

**Figure 2.  Distribution of the Longest Run of Field Goals in Halves: Predicted vs. Observed**

| Length | #Predicted | #Observed |
|--------|-----------|-----------|
| 2 | 30.2 | 19 |
| 3 | 556.8 | 527 |
| 4 | 905.7 | 928 |
| 5 | 562.1 | 573 |
| 6 | 248.5 | 248 |
| 7 | 97.8 | 106 |
| 8 | 37.0 | 43 |
| 9 | 13.8 | 10 |
| 10 | 5.1 | 2 |
| 11 | 1.9 | 2 |
| 12 | 0.7 | 2 |
| Mean | 4.41 | 4.44 |

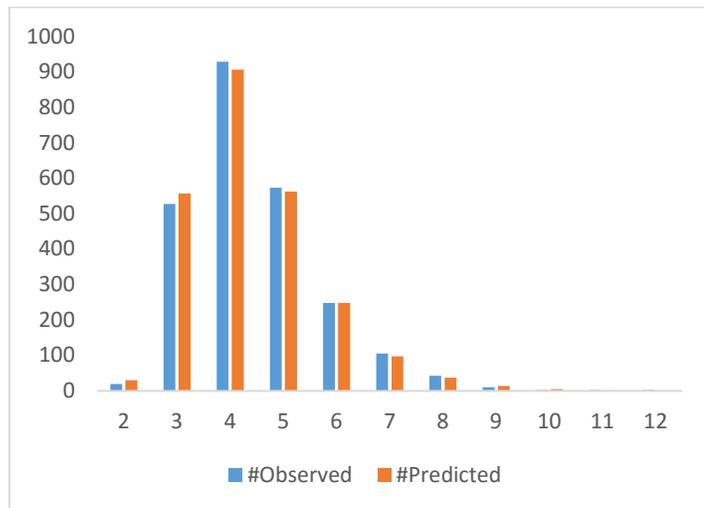

The agreement between the theory and the data is outstanding (chi-square $p = 0.28$). There is no evidence here either of long runs of baskets being due to momentum.

# Incorporating Free Throws

Since the above analyses looked only at the patterns of field goals, we also conducted an analysis that defined a scoring event as either a field goal or an instance of one or more made free throws, excluding those immediately following a basket ("and one's") since possession of the ball does not change prior to those free throws. The results are shown in Figure 3:

**Figure 3.  Distribution of the Longest Run of Scores (Field Goals or Free Throws) in Games: Predicted vs. Observed**

| Length | #Predicted | #Observed |
|--------|-----------|-----------|
| 3 | 66.5 | 76 |
| 4 | 405.2 | 371 |
| 5 | 417.4 | 454 |
| 6 | 211.2 | 201 |
| 7 | 83.2 | 93 |
| 8 | 30.1 | 26 |
| 9 | 10.6 | 6 |
| 10 | 3.7 | 2 |
| 11 | 1.3 | 1 |
| Mean | 5.00 | 4.99 |

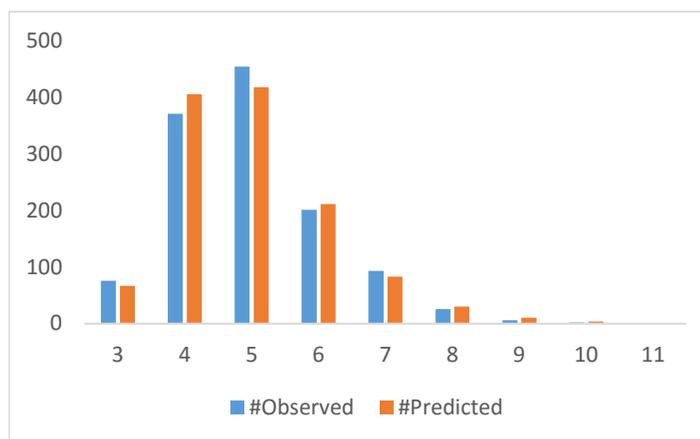

Once more we see no significant discrepancy between the actual lengths of the longest runs of scores and what probability theory predicts in the absence of momentum ($p = 0.07$).

# Accounting for Differences in Team Abilities

In each of the above investigations we have treated the likelihood that the same team that just scored scores the next basket is equal for each team. However within a given game this probability could be measurably higher for one team than the other—for example when the Golden State hosts a last place team, we would expect the chance of scoring consecutively to be greater for the Warriors than for their opponent. It is known that when success probabilities vary, analyses using the mean success probability typically give very similar results to those that derive from using the actual success probabilities (Hodges and Lehmann, 1970). Just to be sure, however, we did two final analyses involving only those games in which the teams were likely to be fairly evenly matched. Specifically, we just considered games in which the visiting team's point differential for the season was within three points either way of the home team's point differential plus five, the five point addition serving to account for home court advantage. 350 games met this criterion.

Figures 4 and 5 provide the results of this investigation, first using runs of field goals only, then also including free throw scores in the manner specified previously. Restricting attention to games involving closely matched teams yields, as in the previous cases, an excellent fit between the observed longest run lengths and what is expected under the null hypothesis of no momentum ($p = 0.82$ and $p = 0.47$ respectively for the two cases below):

**Figure 4. Distribution of the Longest Run of Field Goals
in Games for Closely Matched Teams: Predicted vs. Observed**

| Length | #Predicted | #Observed |
|---|---|---|
| 3 | 18.11 | 17 |
| 4 | 104.23 | 104 |
| 5 | 115.54 | 109 |
| 6 | 65.101 | 74 |
| 7 | 28.502 | 28 |
| 8 | 11.38 | 14 |
| 9 | 4.3931 | 1 |
| 10 | 1.6749 | 0 |
| 11 | 0.6356 | 2 |
| 12 | 0.2407 | 1 |

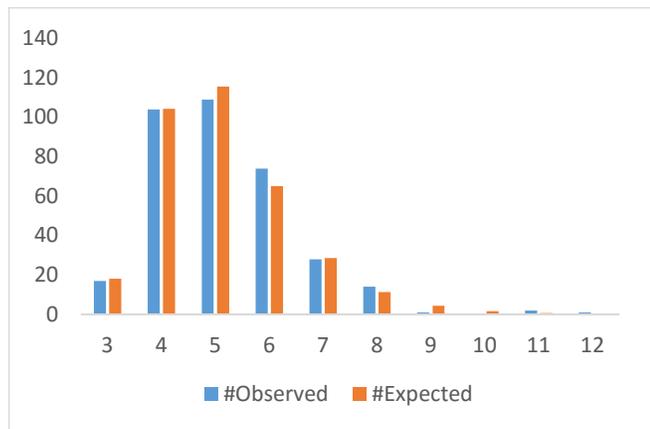

**Figure 5. Distributions of the Longest Run of Scores (Field Goals or Free Throws)
in Games for Closely Matched Teams: Predicted vs. Observed**

| Length | #Predicted | #Observed |
|---|---|---|
| 3 | 19.5 | 28 |
| 4 | 116.2 | 107 |
| 5 | 118.4 | 122 |
| 6 | 59.5 | 64 |
| 7 | 23.4 | 20 |
| 8 | 8.4 | 7 |
| 9 | 3.0 | 2 |

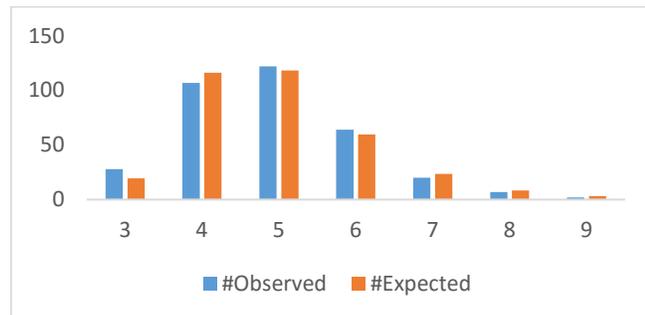

## Conclusion

Whether considering just field goals or including free throw scoring events, we have seen that while there are occasionally runs in NBA games as long as 11 or 12 scores in a row, the lengths of the longest runs in NBA games are entirely consistent with what occurs when scoring events are produced by a random process. Impressions of momentum in team scoring in the NBA appear to be an illusion. We have not investigated run patterns in college basketball; it is still possible that there is detectable evidence of a team "hot hand" (or "cold hand") effect there.

So is NBA basketball "a game of runs?" Not if the implication is that when a team goes on a run, it is because it is "hot"—or because the other team is "cold". NBA basketball IS a game of runs—but only because natural random processes generate many more long runs than most observers expect to see.

# Appendix:  Exact Distribution of the Longest Run

There is more than one way to determine the exact distribution of the length of the longest run of successes in $n$ independent Bernoulli trials with success probability $p$. Schilling (1990) provides a recursive formula that is simple to implement computationally. DasGupta (2018) notes that the following Markov chain model can also be used (also easy to code):

Given positive integer $m$, let $X_k$ be the length of current success run as of the $k$th trial, except that if this length ever reaches $m$ then $X_k$ remains at $m$ from then on (this is known as an *absorbing state*). The possible states of the chain are $\{0, 1, 2, \ldots, m\}$, and the associated *transition matrix* $\boldsymbol{T}$ has entries $T_{i,j} = P(X_{k+1} = j \mid X_k = i)$ given by $T_{i,i+1} = p$ and $T_{i,0} = 1 - p$ for $i = 0, \ldots, m - 1$; $T_{m,m} = 1$. All other entries of $\boldsymbol{T}$ are 0.

From the theory of Markov chains, we know that the $n$-step transition probabilities $P(X_{k+n} = j \mid X_k = i)$ are the elements of the $n$th power of $\boldsymbol{T}$. And from this fact we can find the chance that the length of the longest success run is at least $m$ in the upper right entry of this $\boldsymbol{T^{(n)}}$ matrix, $P(X_n = m \mid X_0 = 0)$. Performing this calculation for all positive integers $m$ determines the exact distribution of the length of the longest success run in $n$ trials.

# References

Albright, S. C. 1993. A statistical analysis of hitting streaks in baseball. *Journal of the American Statistical Association*, 88(424): 1175-1183.

DasGupta, A. 2018. Personal communication.

Ayton, P. and Fischer, G. 2004. The hot hand fallacy and the gambler's fallacy: Two faces of subjective randomness? *Memory & Cognition*, 32(8): 1369-1378.

Crosson, R. and J. Sundali. 2005. The Gambler's Fallacy and the Hot Hand: Empirical Data from Casinos. *The Journal of Risk and Uncertainty* 30(3): 195–209.

Gilovich, T., Vallone, R. and Tversky, A. 1985. The hot hand in basketball: On the misperception of random sequences. *Cognitive Psychology*, 17: 295-314.

Hodges, J.L. and Lehmann, E. L. 1970. *Basic Concepts of Probability and Statistics.* Holden-Day.

Miller, J. B. and A. Sanjurjo. 2015. Surprised by the Gambler's and Hot Hand Fallacies? A Truth in the Law of Small Numbers. *Econometrica* 86(6): 2019-2047.

Schilling, M. F. 1990. The Longest Run of Heads. *The College Mathematics Journal* 21(3): 196-207.

Schilling, M. F. 2012. The Surprising Predictability of Long Runs. *Mathematics Magazine* 85(2): 137-145.

Stern, Hal S. 1997.  Judging Who's Hot and Who's Not. *Chance*, 10: 40-43.

Vergin, R. 2000. Winning Streaks in Sports and the Misperception of Momentum. *Journal of Sport Behavior*, 23: 181-197.